# Does Culture Matter? Impact of Individualism and Uncertainty Avoidance on App Reviews

Ricarda Anna-Lena Fischer
Maastricht University
ra.fischer@alumni.maastrichtuniversity.nl

Rita Walczuch
Maastricht University
r.walczuch@maastrichtuniversity.nl

Emitza Guzman
Vrije Universiteit Amsterdam
e.guzmanortega@vu.nl

*Abstract*—Mobile applications are often used by an international audience and therefore receive a high daily amount of user reviews from various countries. Previous work found evidence that app store reviews contain helpful information for software evolution processes. However, the cultural diversity of the reviews and its consequences on specific user feedback characteristics has only been researched to a limited extent so far. In this paper, we examine the influence of two cultural dimensions, Individualism and Uncertainty Avoidance on user feedback in Apple app store reviews written in different languages. For this purpose, we collected 647,141 reviews from eight countries and written in five languages over a period of six months. We then used manual content analysis and automated processing to examine a sample of 3,120 reviews. The results show that there is a statistically significant influence of Individualism and Uncertainty Avoidance on user feedback characteristics. The results of this study will help researchers and practitioners to reduce algorithm bias caused by less diversified training and test data and to raise awareness of the importance of analyzing diversified user feedback.

## I. INTRODUCTION

Due to the emerging developments in the mobile phone sector, the market for mobile applications is increasingly growing and offers developers opportunities to create low-cost products. However, this growth has also led to a highly competitive market [1]. Thus, it is crucial for developers to consider user feedback when developing and evolving software. In 2019, 1.84 million mobile applications (apps) were available in the Apple app store [2]. Each app in widely used app stores has a review and rating function, making it possible for users to give feedback. Previous work found that app reviews contain valuable information for software evolution, such as bug reports or feature requests [3] and that developers regularly consider this feedback [4].

In large app stores such as the Google Play or Apple app store, about 65% of the apps are downloaded internationally [1]. Software features, as well as app preferences and usage by users from various countries differ [5]. Aligned to this, users may stop using an app if it frequently contains bugs or lacks necessary features [5]. In addition, research found that there are cultural differences among both developers' and users' behavior pertaining to software use and development e.g., [6], [7]. These results raise the question of the *cultural influence* on user feedback. If user feedback is culturally influenced, app developers may rely too much on user feedback from certain cultures and therefore, unknowingly have a competitive disadvantage in cultures that value other software features. If there is no adequate alternative, users only have the choice between using an app that does not meet their requirements or not using such an app at all. Until now, most approaches for automatically processing user feedback e.g., [8], [9], [10], [11], [12] have used reviews stemming from the US and written in the English language. Not considering cultural differences in user feedback could lead to algorithm bias when designing, testing and validating approaches for its automatic processing. Thus, the importance of considering culturally diverse user feedback is essential for users, practitioners and researchers.

Despite the potential influence of cultural differences on user feedback, there are few studies that have investigated this topic. To our best knowledge, only one study exists that has looked at the impact of culture in reviews written in English [13]. However, this work did not examine app reviews written in different languages and only investigated certain cultural dimensions. We address this gap in our current work, which reports on a study analysing cultural differences in reviews written in different languages. For this purpose, we used a well known cultural model [14], that has been widely used in the study of cultural differences in the software engineering context [15]. In our work, we focus on two cultural dimensions: Individualism and Uncertainty Avoidance. In our previous work [13] on cultural differences on app reviews, we did not investigate this point, as only countries with similar Uncertainty Avoidance scores were selected. Our current work sheds some light on this aspect by focusing on the analysis of countries with different Individualism and Uncertainty Avoidance scores.

For the analysis, we extracted 647,141 reviews from eight different countries and 15 mobile applications in five languages (Chinese, English, French, German and Spanish), written in a six month period. We manually analyzed a sample of 3,120 reviews and then further investigated the results through statistical tests[1].

This work contributes to the cultural research of user feedback for software evolution processes in several ways. To our best knowledge, we are the first to examine cul-tural differences (in particular Individualism and Uncertainty Avoidance) in user feedback characteristics in app reviews from various languages. Moreover, our study is the first to investigate the impact of the Uncertainty Avoidance cultural dimension on user feedback characteristics. Furthermore, we examine cultural differences in user feedback characteristics from unpaid and paid apps and thus, broaden the knowledge in this field. Lastly, we are the first to study cultural differences in user feedback characteristics in Apple app reviews for apps that are not only headquartered in the USA.

---

[1]Our replication package can be found here: https://doi.org/10.6084/m9.figshare.13864295

This study can help software developers to better understand their gathered user feedback and adjust their collection methods to a global distributed audience. Furthermore, it raises awareness of cultural differences in user feedback and the resulting potential bias in algorithms that were designed, tested and validated with Apple app store reviews.

## II. BACKGROUND WORK

### A. Software Evolution

Prior work found evidence that app store reviews contain valuable information for software evolution, such as feature requests or bug reports [3]. Subsequently, several researchers developed approaches to classify, summarize and rank app reviews but used mostly English written reviews for this purpose e.g., [8], [9], [16]. Few research has investigated differences in reviews from various countries. Previous work found, for example, that users from Australia, Canada and Japan prefer not to rate apps, while Chinese users rate apps more likely than users from different countries [5]. Additionally, existing research used a random forest algorithm to find factors that differentiate app reviews from the US and other countries [17]. While some work proposed approaches for the automatic processing of reviews written in different languages, e.g., [18], [19] there is relatively few work in this direction. Guzman et al. [13] first investigated the question of cultural differences regarding user feedback in app store reviews. Focusing on eight countries and seven applications, this work found correlations between different cultural dimensions— Indulgence, Power Distance, Individualism— and various characteristics of reviews (sentiment, text length, rating, gender and content) in unpaid mobile applications. However, they only chose countries that have similar Uncertainty Avoidance indexes as well as reviews only written in English. We address this research gap in our current work and thus extend our previous work.

### B. Culture

Culture is often seen as "the collective programming of the mind which distinguishes the members of one group or society from those of another" [20]. In our work we use Hofstede's well known cultural model [14] to analyze cultural differences in Apple app reviews. He surveyed employees of a big company operating in 40 countries. Based on the employees' answers, Hofstede constructed several cultural dimensions. Every analysed country got a specific score for each cultural dimension assigned [14]. Within the scope of this study, Hofstede discovered that there are four dimensions— later, six dimensions— with which cultural differences can be explained. Although other cultural models exist, we assume that Hofstede's cultural model fits the most since it has been validated many times and focuses on cultural values that can potentially influence users' actions (see [21] for examples). Also it is the most widely used model in software engineering contexts [15]. Hofstede's six dimensions are [22]:

- **Power Distance:** the degree to which people in a society accept that power is not equally distributed.
- **Uncertainty Avoidance:** the degree to which members of a specific culture desire to know things that might occur in the future.
- **Individualism/Collectivism:** the degree to which individuals take care of themselves and a group of people.
- **Masculinity/Femininity:** the degree to which values that are assigned to the male or female gender are exhibited in a culture.
- **Long-/Short-term Orientation:** the degree to which someone considers the distant future or only the near future and recent past.
- **Indulgence:** the degree to which people in society can freely enjoy their lives and acknowledge happiness.

In our study, we focused only on Individualism (IDV) and Uncertainty Avoidance (UA) because Individualism, Uncertainty Avoidance and Power Distance are the most frequently used cultural dimensions in information systems [21]. Among these, Individualism is considered to be the most important cultural dimension for explaining differing human behavior e.g., [23], [24]. Power Distance and Individualism are highly correlated (r=-0.67). Using both dimensions could, potentially, lead to multicollinearity and thus, we did not investigate Power Distance.

Previous research found that Uncertainty Avoidance can play a role in differing user behavior (see [21] for examples). However, previous work [13] on user feedback in app reviews, did not consider this cultural dimension, as only countries with similar Uncertainty Avoidance scores were selected. Therefore, we focus on this dimension in this study.

## III. HYPOTHESES

Building upon existing literature, we developed our research hypotheses using common user feedback characteristics such as text length, rating, sentiment and content categories.

**Text length** Previous work found that people from individualistic cultures tend to be more direct [24], [25]. Furthermore, people from collectivistic cultures do not explain their ideas as much as individualistic cultures because they are assuming that it is clear from the context [26]. This point could be reflected in the length of the reviews. The assumption of varying text length is supported by previous studies on cultural differences in Amazon reviews, e.g., [27]. They found that users from the US (high Individualism) write longer reviews than Chinese users (low Individualism). This is aligned with previous research results that showed a positive correlation of text length with Individualism [13].

> *H1: Users from individualistic cultures give more feedback as measured by the length of the review text compared to users from collectivistic cultures.*

Research has shown that cultures with high Uncertainty Avoidance have higher privacy concerns compared to low Uncertainty Avoidance cultures [28], [29]. This could lead to users from high Uncertainty Avoidance cultures writing less in order to disclose less from their data. Based on these previous findings, we formulate the second hypothesis as follows:

> *H2: Users from high Uncertainty Avoidance cultures give less detailed feedback as measured by length of review text compared to users from low Uncertainty Avoidance cultures.*

**Rating and Sentiment** Previous work found that collectivistic cultures seek to have good relationships with group members [30] and seek to preserve harmony [26]. Other work has shown that individualistic cultures tend to express their emotions more often in reviews [31], [32]. Moreover, they discovered that individualistic cultures deviated more from previously given reviews which could also have an impact on the review rating and sentiment by making them more diverse [31]. Previous work examined Booking.com[2] reviews and found that individualistic cultures give fewer positive ratings than users from collectivistic cultures [33]. Based on these findings, we hypothesize that feedback from collectivistic cultures is more positive in terms of rating and contains less emotions. Since emotions in reviews can, potentially, be expressed in a high fluctuation in ratings as well as sentiment, we formulate three hypotheses in this respect:

> *H3a: The review sentiments from users from individualistic cultures fluctuate more than from users from collectivistic cultures.*
> *H3b: The review ratings from users from individualistic cultures fluctuate more than from users from collectivistic cultures.*
> *H3c: Users from individualistic cultures give lower ratings compared to users from collectivistic cultures.*

People from high Uncertainty Avoidance cultures tend to show less tolerance regarding others, and be more extreme and conservative [26]. Considering these two aspects, reviews from users who come from a high Uncertainty Avoidance culture could report matters in a more extreme and less tolerant way, which might have a negative impact on the rating of the app. For instance, previous work found that Chinese consumers tend to give higher ratings than US consumers since they have a lower Uncertainty Avoidance index [27].

In addition, researchers have already studied online reviews from Booking.com and TripAdvisor[3] and found that users from high Uncertainty Avoidance cultures give lower ratings than users from low Uncertainty Avoidance cultures [33], [34]. Assuming a similar behavior in the context of app reviews, we hypothesize that users from high Uncertainty Avoidance cultures giver lower ratings than users from low Uncertainty Avoidance cultures.

> *H4: Users from high Uncertainty Avoidance cultures give lower ratings compared to users from low Uncertainty Avoidance cultures.*

**Content Category** Previous studies found that people in collectivistic cultures tend to write reviews about product functionalities more frequently [32]. In addition, researchers found evidence that individualistic cultures give more recommendations than collectivistic cultures in the context of online customer reviews [32]. This could mean that individualistic cultures give proportionally more general complaints or praises that are not useful for software developers to evolve their software than collectivistic cultures. Therefore, we hypothesize that individualistic cultures give less useful feedback than collectivistic cultures. This underlying hypothesis is also supported by the findings of previous work [13] that found a negative correlation between Individualism and the feedback usefulness.

> *H5: Users from individualistic cultures give less useful feedback compared to users from collectivistic cultures.*

In this context, we refer to *useful feedback* as those that contains information that can be used to improve the software, i.e., a bug report, feature request or support request.
Uncertainty Avoidance cultures might want to know about things that occur in the future more often and therefore could submit more feedback to influence the evolution of the software. Previous work has already shown that people from high Uncertainty Avoidance cultures prefer to have more control in strategic issues and that they associate control with opportunities [35]. By submitting useful feedback more frequently, users from high Uncertainty Avoidance cultures could guide the software evolution in their desired direction and thus gain more control. In addition, people from high Uncertainty Avoidance cultures have a specific desire to achieve clarity [22]. This assumption is underlined by the fact that previous work found a positive correlation between the Uncertainty Avoidance score and neuroticism [36]. They have shown in other contexts that higher neuroticism is aligned with more frequent feedback and higher feedback quality [37]. Based on these aspects, we hypothesize that users from high Uncertainty Avoidance cultures will submit more useful feedback than users from low Uncertainty Avoidance cultures.

---

[2]https://www.booking.com/index.de.html

[3]https://www.tripadvisor.de/

*H6: Users from cultures with high Uncertainty Avoidance scores give more useful feedback compared to users from cultures with low Uncertainty Avoidance scores.*

## IV. Study Methodology

We tested the formulated hypotheses by conducting a manual and automatic analysis on a sample of multi-lingual reviews, stemming from culturally diverse countries. We use the terms app review, review and user feedback interchangeably for the remainder of this paper.

### A. Analyzed Characteristics

We considered four user feedback characteristics in our analysis. These characteristics have been used as input in previous work for automatically classifying and prioritizing user feedback e.g., [8], [9], [10], [11], [12], [38] and could lead to algorithm bias, in case cultural differences would be present and not taken into account. The considered characteristics are:

- **Text length of the review**: Character count for the review text.
- **Sentiment of the review**: Affect expressed by the user in the review. Can range from very positive, positive, neutral, negative and very negative.
- **Rating of the review**: Users' satisfaction with the respective app. Can range from 1 to 5.
- **Content of the review**: The content category of the review with respect to software evolution: bug report, feature request, support request, noise and other.

### B. Data Collection

We collected user feedback stemming from different countries. We selected these countries based on their previously computed scores for Uncertainty Avoidance and Individualism[4]. To take into account the possibility that Individualism and Uncertainty Avoidance scores slightly change over time, we consider the *tendencies* of the selected countries, instead of their exact scores. If a country score was above 50, we assigned a *high* value to the respective cultural dimension and if it was below 50, we assigned a *low* value to the respective cultural dimension score. The chosen countries had Uncertainty Avoidance and Individualism scores that deviated most from the threshold score of 50. However, some exceptions were made if there was a lack of reviews and/or manual annotators that were fluent in the respective language during the given time frame. We included user feedback from eight different countries: France, Germany, Mexico, Singapore, Pakistan, People's Republic of China, the USA and the UK, and written in five different languages: Chinese, English, French, German and Spanish. We chose two countries for each Individualism and Uncertainty Avoidance cultural dimensional group. Each cultural group is composed by one possible permutation of the Uncertainty Avoidance and Individualism values (high or low). Thus, there are four cultural groups in the sample. The specific scores for each country are shown in Figure 1.

Fig. 1: Hofstede's Scores for Selected Countries

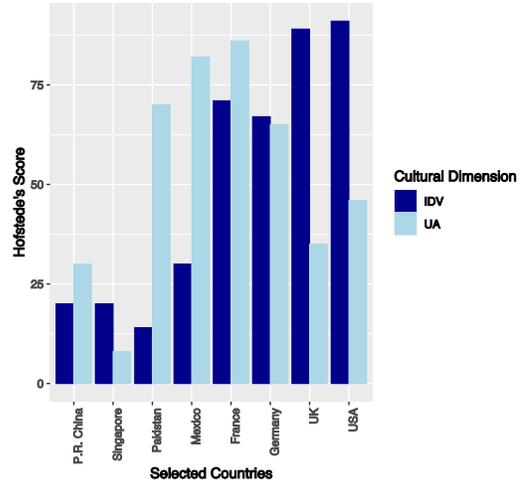

Further, we selected user feedback related to different apps. We chose these apps based on their usage popularity and the high amount of reviews they received in the collection time span. For this purpose, we analyzed the official Apple top app charts[5] and a third-party platform for the countries not included in the official charts[6]. Table 1 shows the 15 selected apps (5 paid/10 unpaid). Eight of the selected apps were from the US, the other apps had their country of origin in Australia, People's Republic of China, Hong Kong, Russia, Sweden and Ukraine; resulting in a diversity in the country of origin of the software that is uncommon in previous work.

The final data collection took place on 18 and 19 March 2020 via a specialized app review collection platform[7]. The final dataset contained 647,141 reviews from eight countries and 15 mobile applications from a time span of six months (10/09/2019-15/03/2020).

### C. Sample Creation

We analysed the reviews characteristics through automatic and manual analysis. For the manual analysis, it was necessary to sample our data, as the consideration of the whole set was unfeasible. Additionally, the amount of reviews differed highly across the analyzed countries, which could have led to a bias in the statistical test and thus, this was another reason to analyze a balanced sample in terms of the amount of reviews per country. To create the sample, we chose a 95 % confidence interval for the minimum sample size. We aimed to equally distribute the reviews per mobile application and country in each cultural sample. Afterwards, we stratified the sample of each app along its rating distribution. The final sample contained 3,120 reviews from the eight countries and 15 mobile applications. Every country in our sample was represented with 390 reviews.

---
[4]https://www.hofstede-insights.com/country-comparison/
[5]https://apps.apple.com/de/story/id1484100916?ign-itsct=BestOfApps
[6]https://sensortower.com/ios/rankings/top/iphone/pakistan/all-categories
[7]https://appfollow.io/

TABLE I: Selected Apps

| Adobe | AutoSleep | FaceApp | Facebook |
|---|---|---|---|
| Forest | Instagram | GoodNotes | Google Maps |
| Notability | Netflix | ScannerPro | Snapchat |
| Spotify | TikTok | YouTube | |

*D. Automatic Content Analysis*

We automatically analyzed the rating and text length of the reviews in our final sample. In contrast to the ratings that could be directly used for the analysis, it was necessary to normalize the text length of the reviews in order to make them comparable across the different countries. This was necessary since the structure of the investigated languages differs [39]. For example, the sentence "I am going home" has 12 characters in English without white spaces, 16 in German, 17 in French, 10 in Spanish and 6 in Chinese. We used the approach of Barbro et al. [39] to normalize the text length of each review. In this approach, they calculated a normalization factor for Chinese, French, German and Spanish based on Amazon product reviews. We used these factors in our normalization and removed emojis and punctuation marks before counting the characters. For its computation, we considered the title and review body.

*E. Manual Content Analysis*

While the rating distribution and text length could be analyzed automatically, a manual content analysis was necessary for the analysis of the content categories and sentiment. For the manual annotation process, ten annotators annotated the sample of 3,120 reviews. To avoid major disagreements between annotators, we used an annotation guide with definitions and examples of the sentiment levels and content categories. The annotation was done through a specialized tool.

We conducted three test trials with 20 - 50 reviews (not part of the final sample) per trial. After each test trial, we modified the annotation guide with additional examples. In the third trial, the mean Cohen's Kappa ranged between 0.63 to 0.79—signaling a substantial agreement between annotators—and we therefore proceeded with the labeling of the final sample.

**Annotation Process** Ten annotators labelled between 190 and 1,950 reviews. Each review was independently annotated by two annotators. For each review annotators labelled the sentiment (very positive to very negative) and content category of each review (bug report, feature request, support request, other or noise). To avoid bias when labelling the sentiment, the tool did not display the ratings. Annotators could only label one content category for each review. The total annotation process took between 64 to 80 hours (the time for the trials is not included here). The Cohen's Kappa ranges from 0.71 to 0.82.

**Manual Disagreement Handling** The disagreement handling between both annotators was conducted manually and automatically. We used an automatic approach for the sentiment and a manual approach for content. If the two annotators disagreed in the sentiment of a specific review, we computed the average among both annotators as the final sentiment score. For doing this, we first converted the categorical scale (very negative to very positive) to an ordinal scale (-2 to +2). This conversion was also used in the statistical analysis detailed in Section V. For the content disagreements, the first author of this work— highly familiar with manual content classification— manually examined the review and selected the content category that she deemed most appropriate for the final annotation. She is fluent in two languages considered in the study. For the remaining three languages, a test-trial with 30 reviews per language was conducted to check whether an automatic translator[8] could translate the reviews in such a way that the content of the reviews was understandable. While this was the case for most of the Spanish and French written reviews, it was not the case for Chinese reviews. Thus, for reviews written in Spanish and French, she used a translator to check the content. If the translation did not make sense and for all Chinese reviews, she asked one of the annotators to directly translate the review and then decided on the appropriate category.

V. RESULTS

**Text Length** The average character length count in our sample is 132.74 (standard deviation: 171.23). After conducting initial statistical tests on homogeneity of variance and normality to identify the text length characteristics, we applied a two-way ANOVA test with a White adjustment to the logarithm transformed converted character counts since this characteristic has non-homogenous variances. Table 2 shows the results for both factors. As can be seen, Uncertainty Avoidance and Individualism are statistically significant (p-value < 0.0001***/ p-value < 0.0001***). However, the results and the interaction plot in Table 2 and Figure 2 reveal that there is no significant interaction effect for the text length characteristic. Thus, the null hypothesis, implying that the sample means are equal, is rejected for the Uncertainty Avoidance and Individualism dimension. The findings show that there is a statistically significant influence of Individualism and Uncertainty Avoidance on the text length of the reviews. Thus, *hypothesis 1*, stating that **users from individualistic cultures give more feedback as measured by the length of the review text compared to users from collectivistic cultures**, is confirmed. Moreover, *hypothesis 2*, stating that **users from high Uncertainty Avoidance cultures give less detailed feedback as measured by length of review text compared to users from low Uncertainty Avoidance cultures**, is confirmed.

**Rating and Sentiment** The average sentiment score in our

---

[8] https://www.deepl.com/de/translator

TABLE II: Test Results for Text Length

|       | DF   | F       | P(>F)      |
|-------|------|---------|------------|
| UA    | 1    | 36.7886 | <0.0001*** |
| IDV   | 1    | 65.2723 | <0.0001*** |
| UA*IDV| 1    | 0.3473  | 0.5557     |
| Residuals | 2996 | | |

p-value < 0.001 '***'; p-value < 0.01 '**'; p-value < 0.05 '*'

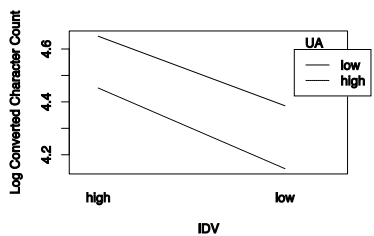

Fig. 2: Interaction Plot Logarithm Transformed Character Count between Uncertainty Avoidance and Individualism

sample is 0.33 (standard deviation: 1.29), indicating an overall neutral sentiment. The average rating is 3.51 (standard deviation: 1.68). As a first step, we tested whether the fluctuations of sentiment were statistically significant. The p-value for the null hypothesis, implying that variances are homogeneous across the Individualism levels, shows that the variances of the sentiment for the Individualism dimension are different with statistical significance (p-value <0.0001***). The descriptive statistics suggest, that the sentiment for individualistic cultures fluctuates more (standard deviation high IDV/high UA: 1.32, standard deviation high IDV/low UA: 1.54) than from collectivistic cultures (standard deviation low IDV/low UA: 1.1, standard deviation low IDV/high UA: 1.24). Thus, *hypothesis 3a*, that **the review sentiments from users from individualistic cultures fluctuate more than from users from collectivistic cultures**, can be confirmed.

Afterwards, we investigated the cultural impact on the rating characteristics. We used a Levene test to analyze whether the ratings of different Individualism levels fluctuate with statistical significance. The result (p-value= 0.209) shows that the null hypothesis stating that the variances are homogeneous cannot be rejected. Therefore, *hypothesis 3b*, that the review ratings from users from individualistic cultures fluctuate more than from users from collectivistic cultures, cannot be confirmed. The test on variance homogeneity indicated that the rating characteristic has homogenous variances, so we used a two-way ANOVA test. The results show that only Individualism has a statistically significant influence on rating (p-value = 0.00311**). For Uncertainty Avoidance, the two-way ANOVA led to a non-significant result (p-value= 0.14350). Furthermore, the results and the interaction plot in Table 3 and Figure 3 reveal that there is no significant interaction effect for the rating characteristic. Therefore, the null hypothesis that all means of the different cultural groups are equal can only be rejected for the Individualism dimension. Thus, *hypothesis 3c*, that **users from individualistic cultures give lower ratings compared to users from collectivistic cultures**, can be confirmed. The findings also suggest that *hypothesis 4*, that users from high Uncertainty Avoidance cultures give lower ratings compared to users from low Uncertainty Avoidance cultures, cannot be confirmed.

TABLE III: Test Results for Ratings

|       | DF   | F     | P(>F)     |
|-------|------|-------|-----------|
| UA    | 1    | 2.141 | 0.14350   |
| IDV   | 1    | 8.753 | 0.00311** |
| UA*IDV| 1    | 0.033 | 0.85593   |
| Residuals | 3116 | | |

p-value < 0.001 '***'; p-value < 0.01 '**'; p-value < 0.05 '*'

Fig. 3: Interaction Plot Rating between Uncertainty Avoidance and Individualism

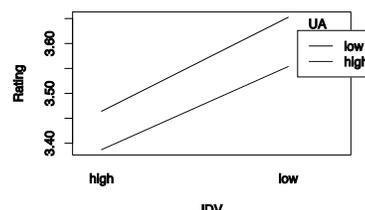

**Content Category** More than half of the analyzed reviews, 53%, contains useful content for software evolution. Roughly 26% are bug reports, 22 % are feature requests, 5 % support requests, 43 % other and 4 % noise. Since the content category is classified with a nominal scale, we conducted a Chi-Square test of independence. The Chi-Square-test results have a significant p-value (p-value < 0.0001***). Thus, the null hypothesis, that no association between the distribution of the content categories and the cultural dimensions exist, can be rejected. Figure 4 shows the content categories share in the sample for each cultural group. To identify whether the share of useful content is influenced by the cultural dimension, we assigned binary variables to the content categories, where 1 stood for useful content for software evolution (bug report, feature request or support request) and 0 for not useful content (other). We classified support request as useful since we assumed that these contain implicit useful feedback for software evolution tasks. For example, if a user asks how to upload a profile picture, the question indicates that either the upload functionality is not easy to find or the provided documentations are not written well enough and thus is an implicit user feedback that can be used to improve the application.

Fig. 4: Content Category Share per Cultural Group

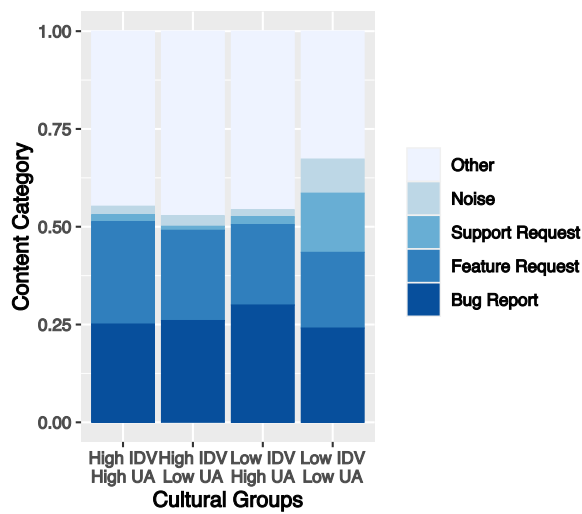

After assigning the binary variables to the reviews in the sample, the Chi-Square test was run again to see if this made a difference. Table 4 shows the results for this test. The results show that only Individualism has a statistically significant influence on the amount of useful content for software evolution. Since the Chi-Square test can only identify whether there is a difference between the four cultural groups, we performed a Chi-Square post hoc test, to identify which cultural group combinations have a statistically significant association to useful content. The Bonferroni method was used to correct the problem of multiple comparisons. The result indicates that only the cultural group combination "high Individualism/ low Uncertainty Avoidance" and "low Individualism/ low Uncertainty Avoidance" is statistically significantly different with respect to useful content[9]. Thus, *hypothesis 5*, that users from individualistic cultures give less useful feedback compared to users from collectivistic cultures, can only be partly confirmed. In other words, **users from individualistic cultures give less useful feedback when they also come from a culture with low Uncertainty Avoidance**. *Hypothesis 6*, that users from cultures with high Uncertainty Avoidance scores give more useful feedback compared to users from cultures with low Uncertainty Avoidance scores, cannot be confirmed. Table 5 summarizes the findings of this study.

TABLE IV: Test Results for Useful Content

|  | $X^2$ | DF | P-value |
|---|---|---|---|
| UA | 0.68172 | 1 | 0.409 |
| IDV | 4.7952 | 1 | 0.02854* |
| Cultural Groups | 12.011 | 3 | 0.007347** |

p-value < 0.001 '***'; p-value < 0.01 '**'; p-value < 0.05 '*'

## VI. DISCUSSION

### A. Findings

Do Individualism and Uncertainty Avoidance have an impact on user feedback characteristics in Apple app store reviews? Based on the results of our study, we can say that in general, yes. The study found evidence that Individualism has a statistically significant influence on the text length, ratings, sentiment fluctuations and, in combination with low Uncertainty Avoidance, on the content categories of Apple app store reviews.

**Individualism** The effect of Individualism supports the hypothesis that individualistic cultures generally write lengthier reviews and thus confirm the findings of previous work [13]. The significant impact of Individualism on text length could be related to two factors. While collectivistic cultures usually use high-context communication, individualistic cultures tend to use low-context communication [26]. Accordingly, collectivistic cultures tend to assume that people already know the necessary information and therefore do not need to pronounce it, while individualistic cultures tend to write out the information. Furthermore, in individualistic cultures, the expression of honest opinions is appreciated, while in collectivistic cultures the goal is to preserve harmony and to avoid confrontations [26]. This characteristic could also lead to users from collectivistic cultures being more likely to give higher ratings and have less sentiment fluctuations in order to avoid conflicts, while users from individualistic cultures are more likely to express their honest opinions and emotions in the ratings and sentiments. Confirming these assumptions, we found that Individualism has a statistically significant influence on the ratings of reviews. The analysis in the Individualism dimension shows that highly individualistic cultures give less positive ratings and have higher fluctuations in sentiments. However, the ratings of different Individualism levels do not fluctuate with statistical significance.

---

[9]Classifying the support requests as not useful content also provided a statistically significant difference.

[10]p-value < 0.001 '***'; p-value < 0.01 '**'; p-value < 0.05 '*'

TABLE V: Hypothesis and Results Summary

| Number | Hypothesis | Confirmed? |
|---|---|---|
| 1 | Users from individualistic cultures give more feedback as measured by the length of the review text compared to users from collectivistic cultures. | Yes ***[10] |
| 2 | Users from high Uncertainty Avoidance cultures give less detailed feedback as measured by length of review text compared to users from low Uncertainty Avoidance cultures. | Yes *** |
| 3a | The review sentiments from users from individualistic cultures fluctuate more than from users from collectivistic cultures. | Yes *** |
| 3b | The review ratings from users from individualistic cultures fluctuate more than from users from collectivistic cultures. | No |
| 3c | Users from individualistic cultures give lower ratings compared to users from collectivistic cultures. | Yes ** |
| 4 | Users from high Uncertainty Avoidance cultures give lower ratings compared to users from low Uncertainty Avoidance cultures. | No |
| 5 | Users from individualistic cultures give less useful feedback compared to users from collectivistic cultures. | Yes ** but only in a low UA context |
| 6 | Users from cultures with high Uncertainty Avoidance scores give more useful feedback compared to users from cultures with low Uncertainty Avoidance scores. | No |

For the content of the reviews, we could only partly confirm that collectivistic cultures provide more useful feedback (bug report, feature request or support request) than individualistic cultures. In fact, Individualism has only a significant influence on content in combination with low Uncertainty Avoidance. The influence of Individualism on the usefulness of content could be related to the fact that individualistic cultures are more likely to express their opinions and give more frequent recommendations than collectivistic cultures [32]. These aspects could lead to the fact that they publish reviews more frequently on other topics, not necessarily useful to software evolution, such as general complaints or praises.

**Uncertainty Avoidance** For Uncertainty Avoidance, the study found evidence that this dimension has a statistically significant impact on text length. The findings show that high Uncertainty Avoidance cultures write shorter reviews than low Uncertainty Avoidance cultures. One explanation for this finding could be that high Uncertainty Avoidance cultures have higher privacy concerns [28], [29] and thus prefer to write less in order to disclose less from their data.

While Individualism has a partly significant influence on the usefulness of the reviews' contents with respect to software evolution, the results do not confirm the hypothesis that high Uncertainty Avoidance cultures give more useful feedback than low Uncertainty Avoidance cultures. This could be due to the desire to avoid uncertainty in the future and because keeping the status quo could override the preference for having influence and control [26]. Previous findings indicate that ratings from low Uncertainty Avoidance cultures are higher than from high Uncertainty Avoidance cultures [27], [33]. However, the results of this study could not support these findings. The prior research was conducted on product [27] and hotel reviews [33], and it could be the case that the cultural influence of Uncertainty Avoidance on ratings differs per investigated review category.

The results for the usefulness of the review content also show that the overall share of useful feedback in Apple app stores has increased from around 30% [3] to approximately 50% in this study, showing that users use the app store more frequently to report bug reports/ feature requests or ask for support than in previous years.

*B. Implications*

**Theoretical Implications** The results have specific implications for academics and practitioners alike. Researchers frequently use app review characteristics for designing, training, validating and testing algorithms that classify, rank or summarize user feedback for software development and evolution processes e.g., [8], [16], [40]. With few exceptions, e.g., [17], [18], [19], previous work has focused on user reviews from the US, or written in English, to design, validate and test their algorithms. However, this can lead to biases in the algorithms. For example, ratings are sometimes used for ranking reviews [8]. User reviews with lower ratings are ranked higher because it is assumed that the issues in these reviews are more urgent and the users more dissatisfied than in the higher rated reviews [8]. Subsequent work has used this approach e.g., [16]. However, the findings in this study show that users from individualistic cultures give statistically significant lower ratings than users from collectivistic cultures. Using this ranking rationale, issues from collectivistic cultures would tend to be ranked with lower priority, leading to a bias. Thus, critical issues reported in collectivistic cultures could be missed due to the fact that they give in general higher ratings. Other review characteristics are also used as a basis for algorithms processing user feedback. For example, previous studies e.g., [9], [41] have used the sentiment in a review to summarize or classify reviews. However, the findings of this study show that the sentiment of cultures with lower Uncertainty Avoidance is statistically significantly lower than that of cultures with high Uncertainty Avoidance. Since researchers often only use reviews from countries with lower Uncertainty Avoidance, such as the USA or UK, to build their algorithms, this can lead to distortions in the algorithm towards the interpretation of positive and negative sentiment, and thus to an overall biased view on the sentiment of users who have written a review. In addition to sentiment, text length was also used by researchers as a characteristic to classify reviews under the assumption that longer texts are more informative e.g., [12] and are therefore more likely to belong to categories that are useful for software evolution, such as bug reports or feature requests. However, the findings of this study show that there is a cultural difference in the text length of reviews. Users from cultures with low Uncertainty Avoidance or higher Individualism write longer texts. Cultural bias can occur if only (or more frequently) longer texts are classified as informative.

Previous research has automatically processed reviews to classify or rank its content e.g., [8], [12], [40]. The findings of our study show that content differs significantly in different cultural groups. However, reviews from highly individualistic and low Uncertainty Avoidance cultures such as the USA or UK are often used to create the algorithms, which can lead to a bias in the algorithm, as the distribution of review content in this cultural group is different from that of other cultural groups. Accordingly, algorithms may be able to better classify categories that are more common in reviews of frequently used cultures than in the less frequently used cultures. One reason for this is that when training and testing the algorithms, more test data from reviews with the categories distribution of cultures preferred for such tasks were used.

We hope that our findings encourage researchers and practitioners to gather user feedback written in diverse languages and from countries with varying Uncertainty Avoidance and Individualism levels to avoid a bias when designing, validating and testing algorithms for its processing.

**Practical Implications** Many review tracking pages like AppAnnie[11] use app analytic functionalities based on keywords, sentiments or ratings to give companies an indication on how their mobile application is perceived by their users. Using only countries with similar Uncertainty Avoidance and Individualism scores, can lead to cultural biases in the interpretation of the users' perceptions. Companies should choose countries with different Uncertainty Avoidance and Individualism scores in their analysis if they want to get a more accurate idea of how users perceive their app. Previous research has shown that developers examine user feedback in reviews when evolving their software [4]. Moreover, prior work proved that users stop using a mobile application once it has bugs or lacks necessary features [5] which could be damaging to the success of the mobile application in a highly competitive market [1]. Thus, only considering user feedback from certain culture groups, could, potentially, lead to competitive disadvantages. Since apps are often distributed on a global market, it is important to adjust them to a highly diverse audience [1]. The findings of this study demonstrate that cultural differences in user feedback exist and that these differences should be considered when gathering and processing user feedback in Apple app stores to avoid a bias in their software evolution processes towards certain cultural groups.

---
[11]https://www.appannie.com/de/

## VII. THREATS TO VALIDITY

Although great efforts have been made to keep validity throughout the research high, some aspects that can impact the results of the study need to be considered. These threats to validity are shortly discussed below.

**Cultural Model** One of the biggest threats to validity is the usage of Hofstede's cultural model which, despite its popularity, has often been criticized for taking scores based on one company research only [42], having a western view [43], being limited to borders [44] and being outdated [45]. While some points may in fact be true, studies have found that the scores are relatively stable over time [46]. To accommodate for little changes in the scores over time, high/ low categories were used instead of the exact scores for the Uncertainty Avoidance and Individualism dimension. However, to improve the soundness of the cultural studies in the context of user feedback for software evolution processes, it would be interesting to know if and how high the impact of culture on user feedback is when using other cultural models.

**Selected Mobile Apps** Another limitation of our study is the analyzed mobile applications. Despite taking greatest care in its selection, the majority of chosen apps have their origin in the Western world. Previous studies have found that people from collectivistic cultures prefer products from their own country more than people from individualistic cultures, regardless of whether these products are superior or not [47]. If this is also the case when giving user feedback, there could be a bias towards in-group and out-group behavior. Behavior towards in-group products refers to products or software that has a same or similar country of origin as the person that is submitting the review [47]. It would be interesting to explore if there is a statistically significant difference in user feedback in collectivistic cultures depending on whether the app was developed in a country that is assigned to the in-group or out-group.

**Relying on Human Judgement** The analysis of the sentiment and content categories of the reviews is based on human judgement; which can lead to subjective annotations. The underlying research design reduced this threat to validity by having all items annotated by two reviewers, giving concrete definitions and examples in the annotations guideline, and conducting three rounds of trial annotations in which the annotation guide was adjusted.

**Considering Other Control Variables** Statistically significant differences do not necessarily indicate a causal cultural influence. The statistical differences could also stem from other reasons such as geographical distance of the countries' locations, income differences or differences in education. Thus, future research should study the effect size and statistical significance of the cultural influence on user feedback when having several control variables. The addition of these control variables can show how high the actual influence of culture on user feedback is and how much is explained by other varying factors.

**Apple Users and National Culture** This threat is related to the people that write the reviews considered in this study. It could be that only a specific subset of users write reviews. Moreover, Apple iPhones are expensive products that are not accessible to everyone. Therefore, the users writing the reviews considered in our study might not be representative of the national culture of the respective country. We chose Apple store reviews because their API allows to easily collect reviews related to specific countries; this is not possible in other platforms such as Google Play. Future work should be conducted in a variety of channels to see if the results of our study hold.

## VIII. CONCLUSION

We investigated whether culture, in particular Uncertainty Avoidance and Individualism, has a statistically significant impact on user feedback characteristics in Apple app store reviews. For this purpose we analyzed 3,120 reviews of 15 different applications from eight different countries and five different languages. Our results show that Individualism and Uncertainty Avoidance have a statistically significant impact on text length, ratings, sentiment fluctuations and the usefulness of the content. While highly individualistic and low Uncertainty Avoidance cultures write longer reviews, highly individualistic cultures also give lower ratings and have higher fluctuations in sentiment. The collectivistic in combination with low Uncertainty Avoidance cultural group tends to give more useful content. Furthermore, the research shows that the proportions of useful feedback have increased to approximately 50%, and thus give developers valuable opportunities to gather useful feedback from app store reviews. The findings draw attention to the diversity of reviews from various countries as well as languages, and the impact of these diversity in user feedback characteristics. We hope that our results encourage practitioners and researchers to gather user feedback from different cultures. Furthermore, the results can help researchers and practitioners to reduce cultural bias when designing, testing and validating algorithms based on Apple app store reviews.